\documentclass[fleqn,12pt]{wlscirep}

\pdfoutput=1
\usepackage{amssymb}
\usepackage{amsmath}
\usepackage{graphicx,booktabs,array}
\usepackage{url}
\usepackage{xcolor}
\usepackage{setspace}
\title{Jet cocoon signatures in the early spectra of a gamma-ray burst/supernova}

\author[1]{L.~Izzo}
\author[1,2]{A.~de~Ugarte~Postigo}
\author[3]{K.~Maeda}
\author[1]{C.~C.~Th\"{o}ne}
\author[1]{D.~A.~Kann}
\author[4,1,5]{M.~Della~Valle}
\author[6]{A.~Sagu\'es~Carracedo}
\author[7]{M.~J.~Micha{\l}owski}
\author[8,9]{P.~Schady}
\author[10]{S.~Schmidl}
\author[2,11]{J.~Selsing}
\author[12]{R.~L.~C.~Starling}
\author[13]{A.~Suzuki}
\author[1]{K.~Bensch}
\author[8,14]{J.~Bolmer}
\author[15]{S.~Campana}
\author[1]{Z.~Cano}
\author[15]{S.~Covino}
\author[11]{J.~P.~U.~Fynbo}
\author[16]{D.~H.~Hartmann}
\author[11,17]{K.~E.~Heintz}
\author[2]{J.~Hjorth}
\author[18]{J.~Japelj}
\author[7]{K.~Kami\'{n}ski}
\author[18]{L.~Kaper}
\author[19,20]{C.~Kouveliotou}
\author[7]{M.~Kru{\.z}y\'{n}ski}
\author[7]{T.~Kwiatkowski}
\author[2,21]{G.~Leloudas}
\author[22]{A.~J.~Levan}
\author[11,2]{D.~B.~Malesani}
\author[7]{T.~Micha{\l}owski}
\author[23]{S.~Piranomonte}
\author[18]{G.~Pugliese}
\author[24]{A.~Rossi}
\author[25]{R.~S\'anchez-Ram\'irez}
\author[26]{S.~Schulze}
\author[22]{D.~Steeghs}
\author[12]{N.~R.~Tanvir}
\author[22]{K.~Ulaczyk}
\author[27]{S.~D.~Vergani}
\author[22,12]{K.~Wiersema} 

\affil[1]{Instituto de Astrof\'isica de Andaluc\'ia (IAA-CSIC), Glorieta de la Astronom\'ia, s/n, 18008, Granada, Spain}
\affil[2]{Dark Cosmology Centre, Niels Bohr Institute, Juliane Maries Vej 30,  2100 Copenhagen \O, Denmark}
\affil[3]{Department of Astronomy, Kyoto University, Kitashirakawa-Oiwake-cho, Sakyo-ku, Kyoto 606-8502, Japan}
\affil[4]{INAF - Osservatorio Astronomico di Capodimonte, Salita Moiariello 16, I-80131, Napoli, Italy}
\affil[5]{International Center for Relativistic Astrophysics, Piazzale della Repubblica 2, I-65122 Pescara, Italy}
\affil[6]{The Oskar Klein Centre, Physics Department, Stockholm University, Stockholm, Sweden}
\affil[7]{Astronomical Observatory Institute, Faculty of Physics, Adam Mickiewicz University, ul.~S{\l}oneczna 36, 60-286 Pozna{\'n}, Poland}
\affil[8]{Max-Planck-Institut f\"ur Extraterrestrische Physik, Giessenbachstrasse, 85748 Garching, Germany}
\affil[9]{Department of Physics, University of Bath, Claverton Down, Bath, BA2 7AY, UK}
\affil[10]{Th\"uringer Landessternwarte Tautenburg, Sternwarte 5, 07778 Tautenburg, Germany }
\affil[11]{Cosmic Dawn Center (DAWN), Niels Bohr Institute, University of Copenhagen, Lyngbyvej 2, DK-2100 Copenhagen \O, Denmark; DTU-Space, Technical University of Denmark, Elektrovej 327, DK-2800 Kongens Lyngby, Denmark}
\affil[12]{Department of Physics \& Astronomy, University of Leicester, University Road, Leicester LE1 7RH, UK}
\affil[13]{Division of Theoretical Astronomy, National Astronomical Observatory of Japan, National Institutes of Natural Sciences, 2-21-1 Osawa, Mitaka, Tokyo 181-8588, Japan}
\affil[14]{European Southern Observatory, Alonso de C\'{o}rdova 3107, Vitacura, Casilla 19001, Santiago 19, Chile}
\affil[15]{INAF - Osservatorio astronomico di Brera, Via E. Bianchi 46, I-23807 Merate (LC), Italy}
\affil[16]{Department of Physics and Astronomy, Clemson University, SC 29634-0978, USA}
\affil[17]{Centre for Astrophysics and Cosmology, Science Institute, University of Iceland, Dunhagi 5, 107 Reykjav\'ik, Iceland}
\affil[18]{Astronomical Institute Anton Pannekoek, University of Amsterdam, 1090 GE Amsterdam, The Netherlands}
\affil[19]{Department of Physics, The George Washington University, Washington, DC 20052, USA}
\affil[20]{Astronomy, Physics and Statistics Institute of Sciences (APSIS), The George Washington University, Washington, DC 20052, USA}
\affil[21]{DTU Space, National Space Institute, Technical University of Denmark, Elektrovej 327, 2800 Kongens Lyngby, Denmark}
\affil[22]{Department of Physics, University of Warwick, Gibbet Hill Road, Coventry CV4 7AL, UK}
\affil[23]{INAF - Osservatorio Astronomico di Roma, Via Frascati 33, I-00040 Monte Porzio Catone (RM), 00078, Italy}
\affil[24]{INAF - Osservatorio di Astrofisica e Scienza dello Spazio di Bologna, via Piero Gobetti 93/3, 40129 Bologna, Italy} 
\affil[25]{INAF - Istituto di Astrofisica e Planetologia Spaziali, Via Fosso del Cavaliere 100, I-00133 Roma, Italy.}
\affil[26]{Department of Particle Physics and Astrophysics, Weizmann Institute of Science, 234 Herzl Street, Rehovot, 761000, Israel}
\affil[27]{GEPI, Observatoire de Paris, PSL University, CNRS,  5 Place Jules Janssen, 92190 Meudon, France}

\begin{abstract}
\textbf{Long gamma-ray bursts mark the death of massive stars, as revealed by their association with energetic broad-lined stripped-envelope supernovae\cite{Galama1998,Hjorth2003}. The scarcity of nearby events and the brightness of the GRB afterglow, dominating the first days of emission, have so far prevented the study of the very early stages of the GRB-SN evolution\cite{Cano2017}. Here we present detailed, multi-epoch spectroscopic observations of SN 2017iuk, associated with GRB 171205A which display features at extremely high expansion velocities of $\sim$ 100,000 km s$^{-1}$ within the first day after the burst\cite{Izzo2017b,deUgarte2017}. These high-velocity components are characterized by chemical abundances different from those observed in the ejecta of SN 2017iuk at later times. Using spectral synthesis models developed for the SN 2017iuk, we explain these early features as originating not from the supernova ejecta, but from a hot cocoon generated by the energy injection of a mildly-relativistic GRB jet expanding into the medium surrounding the progenitor star\cite{Bromberg2011,Harrison2018}. This cocoon becomes rapidly transparent\cite{RamirezRuiz2002} and is outshone by the supernova emission which starts dominating three days after the burst. These results proves that the jet plays an important role not only in powering the GRB event but also its associated supernova.}
\end{abstract}

\begin{document}

\flushbottom
\maketitle
\thispagestyle{empty}

\section*{MAIN TEXT}

On December 5 2017, the Burst Alert Telescope (BAT) onboard the Neil Gehrels {\it Swift} Observatory
triggered on a low-luminosity long gamma-ray burst (GRB) ($L_{iso,\gamma}=3.0\times 10^{47}$ erg s$^{-1}$ in the [15 - 150] keV energy range, $T_{90} = 190.0 \pm 35.0$ s)\cite{Barthelmy2017}. Its afterglow was promptly detected by the X-Ray Telescope (XRT) at a position coincident with the outskirts of a grand-design spiral galaxy, Fig. 1, at a redshift of $z = 0.0368$. We obtained spectroscopy of the afterglow 1.5 hours after the GRB, which exhibited a bright continuum and typical nebular emission lines of H$\alpha$, [N \textsc{II}] $\lambda$6584 and the [S \textsc{II}] $\lambda\lambda$6717/32 doublet from the host galaxy\cite{Izzo2017b}. The host is a massive system\cite{Perley2017b} (${\rm \log} (M^*/M_{\odot}) = 10.1 \pm 0.1$), significantly more massive than typical GRB hosts, which are normally metal-poor, star-forming dwarf galaxies, in particular at low redshift \cite{Vergani2015}. 

The early-time properties and physical diversity of supernovae (SNe) associated with GRBs are still poorly understood, relying on a small number of sources. The proximity of GRB 171205A therefore motivated us to undertake an intensive multi-wavelength photometric and spectroscopic follow-up campaign. The light curve shows an unusual behavior and colour evolution of the optical and ultraviolet emission at very early phases, as compared to the rapid decay observed in X-rays. A few minutes after the burst, a first light-curve bump emerged characterizing the emission during the first two days, with fast evolution from UV to redder wavelengths, as shown in Fig. 1. After the second day, the luminosity of the underlying supernova SN 2017iuk started to increase, reaching its maximum $B$-band magnitude on December 16.4 UT, $\sim11.0$ days after the GRB discovery. The absolute $B$ and $V$ magnitudes at peak are $M_B=-17.5\pm0.1$ mag and $M_V=-18.4\pm0.1$ mag, assuming a distance of $D=163$ Mpc, placing SN 2017iuk at the faint end of the luminosity distribution of SNe associated with GRBs\cite{Cano2017}.

We obtained spectroscopic observations at a nearly daily cadence (Fig. 2). The subtraction of the faint afterglow contribution in the optical spectra (see Methods) confirms the presence of the aforementioned additional emission component during the first three days after the burst. This component is already visible in the first spectrum, $\sim1.5$ hours after the GRB discovery, as an excess of flux at blue wavelengths increasing towards the UV. The multi-frequency spectral energy distributions (SEDs), from near infrared (NIR) to X-ray, at multiple epochs during the first two days are shown in Fig. 1. We model this component with a black-body (see Fig. 1 and Methods), motivated by the detection of a thermal component with a temperature of $kT= 86^{+13}_{-9} $ eV and an emitting radius of $r_{BB} = (1.5\pm0.1) \times 10^{12}$ cm in the very early {\it Swift}-XRT emission from $\sim150$ to 400 s after the GRB trigger\cite{Campana2017} (Fig. 3), and the observation of a thermal component in other GRB-SNe\cite{Campana2006,Starling2012}.

The spectrum obtained at Day 0.975 shows broad absorption features, which we identify as Si{\sc II} $\lambda$6355 and the Ca{\sc II} triplet (centered at 8498 \AA\, rest-frame). The expansion velocities, obtained from the minimum of the corresponding broad absorption line, are unprecedented with values of $\sim 100,000$ km s$^{-1}$ and 95,000 km s$^{-1}$, respectively, and a full-width at half-maximum (FWHM) for the Ca{\sc II} feature of $\sim35,000$ km s$^{-1}$. They become more prominent in the spectrum taken on the second day and show rapid evolution in their profiles, from which we derive velocities of 65,000 km s$^{-1}$ for Ca{\sc II} and of 85,000 km s$^{-1}$ for Si{\sc II}, which hint at the emergence of SN 2017iuk\cite{deUgarte2017} associated with this burst. The measured expansion velocities are greater than the fastest velocities ever observed in any known SN type.  For example, maximum velocities up to $\sim40,000-50,000$ km s$^{-1}$ are observed a few days after the burst in most SNe associated with GRBs\cite{Galama1998,Hjorth2003,Modjaz2006,Bufano2012,Xu2013}. These velocities rapidly decline such that at the GRB-SN peak the expansion velocity is normally\cite{Modjaz2016} at  $\sim 20,000-25,000$ km s$^{-1}$. Therefore the non-detection of high-velocity features in all GRB-SNe may well be the result of having missed spectroscopic observations at very early stages of the GRB-SN event. At later stages, additional features become apparent: the C{\sc II} $\lambda$6580 line and a broad blend of Mg{\sc II} $\lambda$4481 and Fe{\sc II} multiplets 27, 37 and 38 are observed at the same blue-shifted velocity of Si{\sc II} $\lambda$6355. On Day 21.027 we also note the emergence of another broad absorption line in the NIR, likely due to He{\sc I} $\lambda$10830.
 
The intense spectral coverage in the early phases of SN 2017iuk shows that a large amount of high-velocity material is required to reproduce the observed features and their evolution. The emission of GRB-SNe is usually reproduced with spectral synthesis models adopting the canonical CO138 model\cite{Iwamoto1998,Nakamura2001}, but introducing a flat density distribution at high velocities. This is different from the behavior of a canonical spherically-symmetric SN explosion model, where the density gradient would steepen towards the outer layers\cite{Matzner1999}. In the case of SN 2017iuk, we developed spectral synthesis models (see Methods) that reproduce the high-velocity components and their evolution, assuming {\it i)} a single power-law density distribution $\rho \propto v^{\alpha}$ with index $\alpha \sim -6$; {\it ii)} a stratified composition structure, where the relative fractions of different characteristic burning products are allowed to vary to take into account the hydrodynamical mixing in the explosion. The resulting simulated spectra match very well the observed spectral sequence from Day 0.975 to Day 14.936 (Fig. 2). Our model is also able to reproduce the high-velocity Si and Ca absorption features located in the region outside the photosphere, which on Day 0.975 and Day 1.947 are expanding at 59,000 and 53,000 km s$^{-1}$, respectively. For the component above 55,000 km s$^{-1}$, where the enhancement of Fe-peak elements is seen, we find an ejected mass of $M=0.13$ M$_{\odot}$ and a total kinetic energy of $E_{kin}=1.1\times10^{52}$ erg. Using the Ni radioactive-heating model\cite{Arnett1982} for the bolometric SN light curve, we estimate the total SN ejected and Ni mass to be $M_{ej}=4.9\pm0.9$ M$_{\odot}$ and $M_{\rm Ni}=0.18\pm0.01$ M$_{\odot}$ (see Methods). The total SN kinetic energy is $E_K=(2.4\pm0.9)\times10^{52}$ erg, in line with the average value inferred for other GRB-SNe\cite{Cano2017}.

The chemical composition of the high-velocity material is different from the abundance pattern inferred for SN 2017iuk using later-time spectra, but is consistent with the composition predicted by substantial mixing of the explosive material in typical GRB-SN models\cite{Maeda2002,Maeda2003}:  we indeed observe an increasing abundance of Ca, Ti, Cr and $^{56}$Ni toward the lower velocity, while at the highest velocity there is a blob of nearly pure $^{56}$Ni and other Fe-peak elements (Fig. 4). This thorough mixing is expected when the GRB central engine injects high-velocity metal-rich blobs over a time span longer than the time scale of the `shock/jet' breakout. Although this process works irrespectively of the injection mechanism\cite{SuzukiMaeda2018}, the abundances derived for the high-velocity features match those expected in a scenario where the GRB jet is responsible for a considerable fraction of the material in the line-forming regions.

These high-velocity features and their peculiar chemical composition are responsible for the blue emission superimposed om the GRB afterglow evolution, which is only observed within the first hours after the burst, see Fig. 1. The flattening in the ejecta density structure is due to the presence of additional material at high velocities, which in turn implies energy input from the central engine on a time scale longer than the shock-wave breakout time of a typical\cite{Yoon2005} Wolf-Rayet star\cite{Moriya2015} ($\sim$ 10 s). We interpret this fact as a signature of a hot cocoon generated as the jet emerges from the progenitor system\cite{RamirezRuiz2002}: as the jet breaks out, the inner material transported by the jet interacts with the external layers and the medium surrounding the progenitor star\cite{Bromberg2011,Harrison2018,SuzukiMaeda2018}, spreading sideways and giving rise to the observed thermal emission\cite{Nakar2017}. 
This hot material rapidly cools and decelerates from highly-relativistic to sub-relativistic velocities, thus explaining the highly blue-shifted spectral features at early epochs. Indeed, hydrodynamic simulations\cite{DeColle2017} of the emission from such a hot cocoon reproduce the luminosity and temperature evolution observed in the early thermal component of GRB\,171205A observed in X-rays. Our observations and analysis show that the outermost edge ejecta of GRB171205A/SN2017iuk is formed by metal-rich material ejected at 0.3$c$ by the jet. On the other hand, the velocities of the ejecta measured at the peak luminosity of SN2017iuk are very similar to those observed in all GRB-SNe at the same epoch, then implying that GRB/SN explosions are very likely jet-cocoon driven.

\newcommand{\araa}{ARA\&A}   \newcommand{\aap}{Astron. Astrophys.}
\newcommand{\aj}{Astron. J.}         \newcommand{\apj}{Astrophys. J.}
\newcommand{\apjl}{Astrophys. J.}      \newcommand{\apjs}{Astrophys. J. Supp.}
\newcommand{\mnras}{Mon. Not. R. Astron. Soc.}   \newcommand{\nat}{Nature}
\newcommand{\pasj}{Publ. Astron. Soc. Japan}     \newcommand{\pasp}{Publ. Astron. Soc. Pac.}
\newcommand{\procspie}{Proc.\ SPIE} \newcommand{\physrep}{Phys. Rep.}
\newcommand{\apss}{APSS}
\newcommand{\solphys}{Sol. Phys.}
\newcommand{\actaa}{Acta Astronom}
\newcommand{\aaps}{Astron. Astrophys. Supp.}
\newcommand{\iaucirc}{IAU Circular}
\newcommand{\prd}{Phys. Rev. D}

\section*{Acknowledgements}

LI acknowledges support from funding associated to Juan de la Cierva Integraci\'on fellowship  IJCI-2016-30940. LI, AdUP, CCT and DAK acknowledge support from the Spanish research project AYA2017-89384-P. AdUP acknowledges support from funding associated to Ram\'on y Cajal fellowship RyC-2012-09975. CCT acknowledges support from funding associated to Ram\'on y Cajal fellowship RyC-2012-09984. DAK also acknowledges support from funding associated to Juan de la Cierva Integraci\'on fellowship IJCI-2015-26153.  KM acknowledges support by JSPS Kakenhi grants (18H05223, 18H04585 and 17H02864). SS acknowledges support by grant DFG Klose 766/16-3 and discussions with Sylvio Klose. RLCS acknowledges funding from STFC. MJM acknowledges the support of the National Science Centre, Poland through the POLONEZ grant 2015/19/P/ST9/04010; this project has received funding from the European Union's Horizon 2020 research and innovation programme under the Marie Sk{\l}odowska-Curie grant agreement No. 665778. RS-R acknowledges support from ASI (Italian Space Agency) through the Contract n. 2015-046-R.0 and from the European Union Horizon 2020 Programme under the AHEAD project (grant agreement n. 654215). The Cosmic Dawn Center is funded by the DNRF. JH was supported by a VILLUM FONDEN Investigator grant (project number 16599). KEH acknowledges support by a Project Grant (162948--051) from The Icelandic Research fund. JJ and LK acknowledge support from NOVA and NWO-FAPESP grant for advanced instrumentation in astronomy.

\section*{Author contributions statement}
LI, KM, AdUP, DAK, MDV, PS, and CCT wrote the manuscript. LI, DAK and AdUP coordinated the follow-up efforts. LI: Main coordination. X-ray and optical data reduction, spectral analysis and SED interpretation. AdUP: GTC spectroscopic data reduction and analysis, discovery of the emerging supernova and the high-velocity components. KM and AS: spectral synthesis modeling and interpretation. ASC: SN data analysis and interpretation. NRT, CCT and DAK: PIs of the VLT and GTC Afterglow/GRB-SN proposals with which all the spectra were obtained. MJM, TM, KK, TK and MK planned and analysed the RBT/PST2 observations. JS and JJ: VLT data reduction and analysis. KEH and DBM led and planned the NOT observations. PS and SS1 contributed with the UVOT and GROND data analysis and interpretation. RLCS contributed to X-ray data analysis and interpretation. DS, KU, and RLCS planned and analysed the GOTO observations. LI, AdUP, DAK, CCT, MDV, JB, SC1, ZC, SC2, JPUF, DHH, KEH, JH, LK, CK, GL, AJL, DBM, GP, SP, AR, RSR, SS2, DS, NRT, SV and KW contributed to observation strategy and planning for X-shooter observations. All authors contributed to the discussion and presentation of the results and reviewed the manuscript. 

\section*{ Additional information}
\footnotesize

\noindent {\bf Correspondence and requests for materials} should be addressed to LI at \texttt{izzo@iaa.es}. \\

\section*{\small Competing interests}
\footnotesize
\noindent The authors declare no competing financial interests. \\



\newpage

\section*{Methods}
\vspace{0.5cm}
\subsection*{Observations} 
\vspace{0.5cm}
\paragraph*{Photometry}

After the discovery of GRB 171205A by the Neil Gehrels {\it Swift} Observatory satellite mission\cite{Gehrels2004}, we promptly started a monitoring program using several telescopes distributed world-wide. We obtained optical and NIR observations in the $g^\prime r^\prime i^\prime z^\prime JHK$ filters with the Gamma-Ray burst Optical and Near-infrared Detector (GROND)\cite{Greiner2008} mounted on the MPG 2.2 m telescope at ESO La Silla Observatory, Chile; additional $V$ and $R_C$ imaging was obtained with STANcam mounted at the 2.5 m Nordic Optical Telescope (NOT) in La Palma, Spain. We also obtained $BVRI$ imaging at early epochs using the 1.5 meter telescope located at the Observatorio de Sierra Nevada (OSN), located in Dilar, Spain, and additional $R_C$ imaging at smaller remote telescopes\footnote{http://itelescope.net},\footnote{http://acgo.it/oa}. We also obtained $BVR_C I_C $ photometry with the Roman Baranowski Telescope
(RBT/PST2), a component of the Global Astrophysical Telescope System
- a pair of high duty-cycle spectroscopic
telescopes\footnote{www.astro.amu.edu.pl/GATS}. The RBT/PST2 is a Planewave 0.7m fully robotic direct-drive telescope located in the Winer Observatory in Arizona, USA. Photometry of the object was performed with an aperture radius of 2 arcsec and aperture correction was applied. The host-galaxy contribution was estimated by using the same aperture on different parts of the galaxy with a  similar brightness profile to the source location.
\\
Additional data were obtained with the Gravitational wave Optical Transient Observer\footnote{\url{https://goto-observatory.org}} (GOTO), an array of wide-field optical telescopes on a common mount, stationed at the La Palma Observatory, Canary Islands, Spain. Observations of GRB\,171205A were carried out during commissioning of the facility, when three telescopes were active. Each GOTO telescope consists of a $\mathrm{D}=40$ cm f/2.5 optical system using a Newtonian-Wynne configuration. Exposures were obtained at eight epochs in groups of $3\times120$ s, cycling through $R$, $G$ and $B$ Baader filters, which were calibrated by cross-matching with APASS DR9, using $r$ (AB), $V$ (Vega), and $B$ (Vega) magnitudes. Each photometric point was created from a stack of images collected during the given night. Photometry was performed on the object with an aperture radius of three pixels and aperture correction was applied. The host-galaxy contribution was estimated by using the same aperture on different parts of the galaxy with a similar brightness profile to the source location.

The entire dataset of our photometric follow-up is presented in Tables 1-8. The values are neither corrected for Galactic extinction along the line-of-sight of $E(B-V)=0.05$ mag nor the one intrinsic to the host with a value of $E(B-V)_{int}=0.02$ mag (see the {\it Data Analysis} section on the derivation of these values).

\paragraph*{Spectroscopy}

After the discovery of GRB\,171205A with {\it Swift}, we immediately activated our programme (Stargate Consortium, PI: N. Tanvir) at the ESO Very Large Telescope (VLT). We observed the bright ($r'\sim16.0$) afterglow just 1.5 hours after the GRB discovery, identifying typical nebular emission lines of H$\alpha$, N{\sc II} $\lambda$6584 and the doublet of S{\sc II} $\lambda\lambda$6717/32 at a common redshift\cite{Izzo2017b} of $z=0.0368$, which confirms the association of GRB\,171205A with the galaxy 6dFGS gJ110939.7-123512.\\
In parallel, we activated our GRB/SN monitoring programmes at the Gran Telescopio Canarias (GTC) with OSIRIS (PIs: C. Th\"one, D. A. Kann). The first spectrum with GTC/OSIRIS was obtained 0.975 days after the GRB, followed by observations with an almost daily cadence and complemented by additional X-Shooter observations. All spectroscopic observations are presented in Table 10.

\paragraph*{High-energy observations}

We also reduced and analyzed X-ray and UVOT data of GRB\,171205A obtained with \textit{Swift}. The XRT data were obtained from the public \textit{Swift} data archive and then reduced with the standard HEASOFT \texttt{XRTPIPELINE} tool (version 6.22.1), after which we extracted time-resolved spectra at six different epochs using \texttt{XSELECT}, see also Fig. 6. Ancillary response files were consequently built with the \texttt{xrtmkarf} tool, using exposure maps for each single observation epoch. As a counter-check for our data reduction, we used the online tool for time-sliced spectra extraction available on the UK \textit{Swift} Science Data Centre\footnote{http://www.swift.ac.uk/}. Both methods give very similar results.

Pipeline-processed UV and optical sky images taken with the {\it Swift} UV and optical telescope (UVOT) \cite{Roming2005} were downloaded from the UK \textit{Swift} Science Data Centre. Individual exposures at any given observing epoch were combined for each of the seven UVOT filters ($v$, $b$, $u$, $uvw1$, $uvm2$, $uvw2$ and $white$). UVOT photometry was obtained following the standard procedure\cite{Poole2008} using a 3 arcsec radius source-extraction region centered at the GRB/SN position. An aperture correction was applied in order to remain compatible with the effective area calibrations, which are based on 5 arcsec aperture photometry\cite{Poole2008}. The background was estimated using a 5 arcsec radius extraction region located at a similar distance from the host galaxy centre as the GRB. In the absence of pre-explosion host-galaxy UVOT template images, we considered our choice of a background region the best way to remove the contamination by host-galaxy light from our GRB/SN photometry. \\

\subsection*{Data analysis}
\vspace{0.5cm}
\paragraph*{The initial two days}  

According to the standard GRB afterglow scenario\cite{Sari1998}, the spectral energy distribution (SED) is the result of synchrotron emission from shock-accelerated electrons, which is best modelled with a power-law function from X-rays to optical frequencies. This implies that extinction-corrected colour indices are generally constant during the entire duration of the GRB afterglow\cite{Simon2001}. At very early epochs, optical data do not show the typical decay observed in GRB afterglows, a behavior that instead is observed in the UVOT $uvw2$, $uvm2$ and $uvw1$ filters from the onset of the GRB. Similar to the evolution observed in the UV filters, a decay phase is also observed for the other UVOT filters ($u$, $b$, and $v$), but starting at later times. This is similar to the observed behavior at early times\cite{Campana2006} in GRB 060218, suggesting the presence of an additional component in the early light-curve evolution. At later times the emerging SN is visible at almost all wavelengths: the SN emission starts to emerge at  $\sim2$ days after the GRB trigger\cite{deUgarte2017}. Evidence for cooling thermal emission peaking at UV and optical wavelengths can be found in the colour index evolution at very early times. We use UVOT data to build colour evolution curves of $u-b$, $b-v$, $uvw1-u$, and $uvw2-uvw1$ (see Fig. 5). Our results show a variation during the first day of the GRB emission for all colour indices except the $uvw2-uvw1$. Late-time variations of the $u-b$ colour index are due to the emerging SN component.  Variations from a constant behavior are expected when we observe the passage of the synchrotron break frequency at optical frequencies\cite{Granot1999}. In our case, mm-observations and spectral indices derived from X-ray spectra suggest that the cooling break frequency lies at longer wavelengths (see next section). Furthermore, the late-time behavior is dominated by the SN emission: we observe a steep rise at optical wavelengths while UV colours remain almost constant, likely due to the effect of line blanketing and some residual emission from the host galaxy (that is visible also at late epochs in the light curve in Fig. 1). 

To further investigate the origin of the first emission bump and the observed colour index variations we constructed six SEDs using our own dataset for the NIR-to-NUV range and \textit{Swift} XRT data for the X-ray part. The SEDs are computed at mean epochs of $T_{SED1}=0.003$ days, $T_{SED2}=0.06$ days, $T_{SED3}=0.17$ days, $T_{SED4}=0.55$ days, $T_{SED5}=0.97$ days and $T_{SED6}=1.95$ days from the GRB trigger. We note that the last two SEDs correspond to the second and third epoch of our spectral dataset. We fit the SEDs using the \texttt{XSPEC} software\cite{Arnaud1996}. We follow the general fitting procedure\cite{Schady2010} where a first absorption component for the Galactic interstellar extinction is considered with\cite{Schlegel1998,Willingale2013} $E(B-V)=0.05$ mag and $N(H\textsc{i}) =5.89\times10^{20}$ cm$^{-2}$, obtained using the Cardelli extinction curve\cite{Cardelli1989}. A second absorption component corrects for the host-galaxy extinction by dust scattering using the Milky Way ($R_V = 3.08$) template\cite{Pei1992}. We estimate the intrinsic $E(B-V)$ value from the observed Na{\sc I} $\lambda$5890 line in the X-Shooter spectra. We use the spectra from Day 7 and Day 21 to measure its equivalent width and obtain an extinction of $E(B-V)_{int}=0.02$ mag according to the Poznanski et al. relation\cite{Poznanski2012}. This value is consistent with the extinction inferred from the Balmer decrement  determined from late nebular spectra (Kann et al. in preparation). From an X-Shooter spectrum at 56 days after the GRB we measure H$\alpha$/H$\beta=2.89$ corresponding to $E(B-V)_{int}=0.01$ mag, assuming an electron temperature $T_e = 10^4$ K and density $n_e = 10^2$ cm$^{-3}$ for case B recombination\cite{Osterbrock}. For the Galactic and host X-ray absorption component, we used the Tuebingen-Boulder ISM absorption model\cite{Wilms2000}. In order to estimate the intrinsic X-ray column density $N_{\rm H,int}$, including the effects of the metallicity, we fit late-time XRT data $t \div (1.5 \times 10^{4}, 3.0 \times 10^{6}$) s obtaining $N_{\rm H,int} = (9^{+6}_{-5}) \times 10^{20}$ cm$^{-2}$. In our SED fits, we fix the $N_{\rm H,int}$ value to the one determined from late-time X-rays.

Our analysis shows that the broad-band SED cannot be fitted with a single power-law function. At optical-UV frequencies we find a clear excess above the extrapolated power-law best-fit function to the X-ray data alone (see Table 12). We exclude the passage of the synchrotron break frequency at optical wavelengths as the origin of the observed colour index variation. In this case our NIR-to-NUV data should follow a power-law function with spectral index $\beta_X+0.5$, but with a lower flux than the one observed. We model this excess as a blackbody, motivated by the presence of a thermal component in very early XRT Windowed-Timing (WT) spectra from $\sim150$ to 400 s after the GRB trigger\cite{Campana2017}. We fitted the X-ray spectra using XRT-WT data in the interval ($T_0 + 185, T_0 + 385$) s, but taking the $N_{\rm H,int}$ value determined at later times (see above). We confirm the presence of an additional thermal component with an observed temperature of $kT= 86\pm4 $ eV and an emitting radius of $r_{BB} \approx 10^{12}$ cm, see also Tables 12 and 13.

The paucity of data (see Table 12) does not allow for a better constraint of the spectral model parameters for the SEDs at days 0.17 and 0.55, leading to relatively large uncertainties for the blackbody temperature and radius. Despite this we observe an evolution of the thermal component with time. The temperature decreases from $\sim200000$ K to 5000 K from Day 0.06 to Day 2, while the radius of the emitter increases from $2\times10^{12}$ cm to $10^{15}$ cm (see Fig. 3). 

\paragraph*{Evolution from 2 days onwards}

From Day 2 on, the light curve at optical wavelength becomes increasingly dominated by rising emission from the underlying SN 2017iuk. We fit the $BVR_C I_C$ light curves of SN 2017iuk with smoothed spline-interpolation functions to determine the maximum of the SN emission. Our results show that SN 2017iuk reached maximum $B$-band magnitude on December 16.4 UT, at 11.0 days after the GRB discovery, see Fig. 6. The absolute magnitude at the peak applying K-correction is $M_B=-17.5\pm0.1$ mag and $M_V=-18.4\pm0.1$ mag for the $B$- and $V$-band respectively. This suggests that SN 2017iuk is one of the faintest broad-lined SNe connected to GRBs\cite{Cano2017}. The faintness of the SN is consistent with the low luminosity of the GRB\cite{Hjorth2013} as well as with the SN luminosity peak at\cite{Cano2017} Day 11. We also compare the absolute-magnitude light curve of SN 2017iuk with the emission of the other two most nearby GRB-SNe known, SN 1998bw and SN 2006aj\cite{Clocchiatti2011,Ferrero2006}. After correcting for Galactic and intrinsic extinction, we compute absolute magnitudes for Johnson $V$ and Cousins $R$ filters, correcting for the K-correction using a specific colour index method\cite{Chilingarian2010}. Our results are shown in Fig. 7.

Finally, we constructed a bolometric light curve of the SN emission using our $uBVR_C I_C JH$ photometry to estimate the initial mass of Nickel in the total ejecta mass of the SN. To this aim, we have considered the radioactive-heating model\cite{Arnett1982,Valenti2008} considering full trapping gamma-rays where the SN is powered by the decay of Ni and Co into daughter products: in this model, the gamma-rays produced in the decay interact with the SN ejecta, thermalizing it and giving rise to the observed SN emission. Assuming an optical opacity of $\kappa=0.07$ cm$^2$ g$^{-1}$ and an observed photospheric velocity at the peak of $v_{phot} = 22000$ km s$^{-1}$, we have obtained a Ni mass of $M_{\rm Ni}=0.18\pm0.01$ M$_{\odot}$, a total mass ejected in the SN of $M_{\rm ej}=5.4\pm1.4$ M$_{\odot}$ and a kinetic energy of $E_K=(2.6\pm1.1)\times10^{52}$ erg. More details on the formulation of the radioactive-heating model used in this analysis and the bolometric light curve construction method can be found in literature\cite{Cano2017b}.

\paragraph*{The spectral analysis}

Optical spectra of GRB171205A/SN2017iuk have been obtained at VLT and GTC between Day 0.06 and Day 29.926 with an almost daily cadence (see Table 9). In order to study the spectral evolution of the SN, as well as of the early thermal component, we need to subtract the GRB afterglow contribution at each epoch in our spectral dataset, so as to obtain a clean spectrum of the supernova component. To determine the GRB afterglow contribution at optical wavelengths, we have considered the observed X-ray emission, which in GRBs mostly tracks the evolution of the optical afterglow\cite{Nousek2006}. In the case of GRB 171205A, the X-ray afterglow is best modeled by a power-law function with spectral index $\beta=-0.85$. From the standard fireball theory\cite{Granot1999} we know that such a value is only valid for the part of the synchrotron spectrum between the peak of the synchrotron emission and the cooling break. $mm$-observations\cite{Perley2017b} indicate that the spectral peak is well beyond NIR wavelengths (de Ugarte Postigo et al., in preparation), hence we can assume that the afterglow emission in the optical range is just the extension of the power-law we observe in X-rays. The result of this subtraction confirms the presence of an extra emission component in the first two days of GRB emission, in particular in the first spectrum obtained $\sim1.5$ hours after the GRB discovery, where we observe an emission excess at blue wavelengths which extends and even increases into the UV, see Fig. 8. A possible absorption feature is also reported at $\sim3700$ \AA, but the lack of additional data at bluer wavelengths prevent a more detailed analysis on its origin. Starting at Day 0.975 we note the presence of broad spectral lines at very high recession velocities, typical of SNe associated with GRBs. On Day 0.975 we measure velocities of $\sim100,000$ km s$^{-1}$ for Si{\sc II} $\lambda$6355, 95,000 km s$^{-1}$ for the Ca{\sc II} triplet centered at $\lambda$8498 and $\sim90,000$ km s$^{-1}$ for the Fe{\sc II} multiplet 42 centered at $\lambda$5169. These velocities rapidly decrease in the following days reaching typical values observed for GRB-SNe of $\sim23,000$ km s$^{-1}$ at Day 10.952.

From Day 8.905 on we also detect C{\sc II} $\lambda$6580 and a blend of Fe{\sc II} (multiplet 27, 37 and 38) and Mg{\sc II} $\lambda$4481 at the same velocities as Ca{\sc II}. The intensity of the Si{\sc II} $\lambda$6355 line slowly decreases and becomes comparable with the C{\sc II} $\lambda$6580 line. At late epochs, the absorption velocities equally decline and reach values of $\sim13,000-15,000$ km s$^{-1}$. In the spectrum of Day 21.027 we observe a broad absorption feature possibly due to He{\sc I} $\lambda$10830 at a velocity of 13,000 km s$^{-1}$ (see Fig. 9). If this feature is due to He{\sc I} we should also detect the He{\sc I} $\lambda$20580 line, which was observed in SN 1998bw\cite{Patat2001}, but this region of our spectra is affected by telluric absorption features. Likewise, we cannot confirm the detection of any other He{\sc I} line since the transitions at $\lambda\lambda$4471,5876,7065, if present, would be blended with other broad systems in this part of the spectrum. This situation is similar to other broad-lined SNe for which He has been detected\cite{Patat2001,Clocchiatti2001,Mazzali2002}. If we exclude Helium as the origin of this feature, the most plausible alternative is Mg{\sc II} $\lambda$10914 at $\sim15,000$ km s$^{-1}$, suggested by the possible detection of Mg{\sc II} $\lambda$4481 at the same epoch and the same velocities.    

\subsection*{Spectral modeling} 
\vspace{0.5cm}
\paragraph*{The spectra from Day 1 to 15}

In this section, we present our spectral synthesis modeling for SN 2017iuk, focusing on a few selected spectra, at day 1.947, day 7.982, day 10.952 (corresponding to the peak of the SN 2017iuk emission), and day 14.936. We also attempt to model the spectrum at day 0.975, but the uncertainty in the subtraction of the underlying afterglow component does not allow a quantitative analysis, thus this spectrum is addressed only qualitatively. For a specific density and abundance structure given as input parameters, we synthesize a model spectrum using Monte-Carlo (MC) radiation transfer simulations with the TARDIS code\cite{Kerzendorf2014}. TARDIS assumes a sharp photosphere with the luminosity and temperature as input parameters to the simulation, which are iteratively constrained through the spectral modeling. For the treatment of the micro-physics we adopt the following parameters: (1) nebular for ionisation; (2) dilute-local thermodynamic equilibrium (LTE) for excitation; (3) detailed Monte-Carlo estimators for radiative rates; (4) the `macroatom' formalism for line-interaction type.  

Previous studies of GRB-SNe and other broad-lined SNe Ib/c suggest that the density distribution at high velocities is flatter than the one predicted by the standard one-dimensional thermal bomb models. As an example, the model 'CO138E30', originally proposed\cite{Iwamoto1998,Nakamura2001} as a model for SN 1998bw and frequently used in similar studies, has a power-law density index as a function of velocity of $\sim-6$ below 40,000 km s$^{-1}$ and $\sim-9$ at higher velocities. However, in case of SN 1997ef, an index of $\sim-4$ at $\geq25,000$ km s$^{-1}$ had been proposed\cite{Mazzali2000}. A similar structure was considered for SN 2016jca (GRB 161219B) to reproduce the observed high-velocity absorption\cite{Ashall2017} in the Si{\sc II} $\lambda$6355 line with $\sim42,000$ km s$^{-1}$ in the spectrum of Day 5.52. In this latter SN high-velocity material had been reported in the earlier spectrum of Day 3.73, for which they derived photospheric velocities of $\sim 44,000$ km s$^{-1}$ and a high abundance of Fe-peak elements in the ejecta. 

The intense sampling at very early times and high-velocity absorptions in several ions seen in the spectra make SN 2017iuk the best opportunity so far to study the structure of the outermost layer of the GRB-SN. We analysed the density structure based on model CO138E30 as first input model to our spectral synthesis simulations, and reached a conclusion similar to the above-mentioned studies for other GRB-SNe, i.e. a lack of a sufficient amount of material at high velocities to account for the high-velocity absorption features seen in all the phases examined here. As our reference model, we first start with the scaled CO138E30 model, in which the density and velocity scales are changed to fit to the estimate of the ejecta mass and kinetic energy by the simple light-curve and spectral analysis in the previous section. To demonstrate that such an ejecta structure is unable to reproduce the observed spectral sequence, Fig. 10 shows the synthetic spectra for two cases based on this ejecta structure; one assuming a typical C+O core abundance, and the other assuming homogeneous mixing of the explosive elements\cite{Nakamura2001}.  It is clearly evident from the earliest spectra that the model lacks a sufficient amount of element absorbers in the ejecta, even including the Fe-peak elements in the homogeneous mixing. We further test a simple ejecta structure assuming a single power-law index of $-6$ (see below); there is not much improvement for the C+O core composition, but there is substantial change in the spectral features once this is coupled with the extensive mixing of the explosively synthesized elements. In particular, the improvement in the first 2 days under the combination of these simple assumptions (a single power-law plus the homogeneous mixing) is striking, and we hereafter take this model as our reference model. 

The physical origin of such high-velocity material (or flattening in the density structure) has not been clarified yet. The most plausible explanation is a long-lasting energy input from a central engine with the energy injection occurring on time scales substantially longer than the shock-wave breakout within the progenitor star (i.e. up to 10 seconds for a Wolf-Rayet progenitor), leading to a flat density structure\cite{SuzukiMaeda2018}. The resulting density structure is well-represented by a single power-law with index $\sim-6$. From a theoretical point of view, this model represents the central-engine driven explosion scenario (e.g. either the collapsar model\cite{MacFadyen1999} or magnetar formation\cite{Metzger2015}) better than an instantaneous thermal bomb model such as CO138E30. Hence, we adopt a single power-law density with the index of -6 as our fiducial model. 

Our reference model captures the basic feature of the theoretically motivated flat density structure and it is coupled with another prediction from such a model: the extensive mixing of newly synthesized elements into the outer layers. To identify important characteristics in the explosion mechanism, we further divide the ejecta structure into several zones and change the abundance distribution in the input models for the radiative-transfer simulations. 
The default abundance pattern is set to typical mass fractions of a CO layer of a star with $M_{\rm ZAMS}=40 M_{\odot}$ (C 0.01, Mg 0.05, Si 0.005 in mass fraction, with the remaining fraction associated with O) added to the half-solar abundance for the other elements. Additional contributions are considered for (Si, S), (Ca, Ti, Cr), (Fe, Co, Ni and $^{56}$Ni) to allow for mixing of the explosively synthesised elements into the outer layers. These are divided into three groups as shown by the parentheses (e.g., Si and S in the same group), and the relative fractions of the elements in the same group are fixed and unchanged. The ratio of S to Si in this additional contribution is fixed to be 0.5; the ratios of Ti/Ca and Cr/Ca are fixed to 0.1 and 0.33; the ratios of Fe/$^{56}$Ni, Co/$^{56}$Ni, and Ni/$^{56}$Ni (where the numerator is for the stable elements) are fixed to 0.2, 0.005, and 0.25, respectively. Note that these ratios reflect typical theoretical predictions for each burning stage. Therefore, in each zone we have only three free parameters to describe the compositions (i.e., the mass fractions of Si/S, Ca/Ti/Cr and Fe/Co/Ni/$^{56}$Ni), which describe the hydrodynamical mixing of different burning zones. The same structure is used for all the spectral sequence. While the structure below the photosphere is not known and thus the earlier phase modeling does not constrain the deeper part, the material at velocities above the photosphere can participate substantially in forming the absorption lines. 

Fig. 2 shows the model spectra compared to the spectra of SN 2017iuk at the five reference epochs. The model luminosity and photospheric velocity are given in Table 10, and the composition pattern (from the explosive component) is shown in Fig. 4. The model contains a total ejecta mass of $\sim2.5M_\odot$ and kinetic energy of $1.7\times10^{52}$ erg above 10,000 km s$^{-1}$. With this model we also find an ejecta mass of $1.1\times10^{-3} M_{\odot}$ and a kinetic energy of $1.2\times10^{50}$ erg for the high-velocity component above 100,000 km s$^{-1}$. Fig. 2 shows that a good fit can be obtained for the spectral sequence from day 0.975 through day 14.936. Absorption features by several ions at extremely high velocities are observed in the earliest spectra and their identification is confirmed by the spectral synthesis model. It should be noted that our spectra show the existence of such high-velocity sub-relativistic material associated with GRB-SNe directly through the optical absorption features, which represents the first time ever known not only in a GRB-SN, but generally for all SN cases. The outermost velocity of the bulk ejecta material is constrained to be $\sim 110,000$ km s$^{-1}$; the lack of material below this velocity or the existence of material above this velocity should be visible through strong absorption. 

The flatter density structure adopted reproduces the overall spectral evolution as well as the existence of the high-velocity absorption features observed in the earliest spectra at Day 0.975 and Day 1.947. The model also explains consistently the high-velocity absorption features seen in the later epochs at a velocity much higher than the photospheric velocity. 
For the spectral sequence synthesised by the homogeneously-mixed power-law distribution (Fig. 10), there is clearly a  strong absorption by Ca II at Day 8 and thereafter. This is remedied by requiring only the modest overabundance of Ca/Ti/Cr group elements above $\sim 55,000$ km s$^{-1}$, while the strong absorption below $\sim 4,500$\AA\ requires a large abundance of Fe-peak elements exceeding the homogeneous-mixing level. If we keep this high abundance of Fe-peak elements between $\sim30,000$ and $\sim55,000$ km s$^{-1}$, then they will produce too much absorption below $\sim5,000$\AA, and the observed peak at $\sim4,500$\AA\ at Day 8 and thereafter will be smeared out. On the other hand, this peak is well-reproduced by Ti II absorption. We therefore require an increasing abundance of Ca/Ti/Cr in this velocity range, while the Fe-peak elements should not be abundant there. Further below $\sim 30,000$ km s$^{-1}$, we see an increasing abundance of both Fe-peak elements and Ca/Ti/Cr to fit to the spectral evolution at Day 11 and thereafter. 

The inferred characteristic composition structure can be compared to the mass fractions expected in the homogeneous mixing of the explosively synthesised species for the CO138 model (Fig. 4). Interestingly, the overall composition structure is largely consistent with the mass fractions expected from the homogeneous mixing of explosive material within the whole ejecta. The heavier elements tend to increase toward lower velocities, suggesting that the mixing is not complete (while it is already substantial). However, in the highest velocity ($>55,000$ km s$^{-1}$) material this is inverted: the mass fractions of $^{56}$Ni and stable Fe-peak elements is overabundant without enhancement of Ca/Ti/Cr. This analysis suggests that: (1) extensive mixing of the newly synthesised elements and pre-explosion progenitor compositions is required, (2) the degree of mixing is nearly at the level of homogeneous mixing but the layered structure is still kept, and (3) there is further inversion of more `explosive' species in the outer and faster region. This characteristic behavior may be similar to what has been suggested\cite{Ashall2017} for GRB 161219B/SN 2016jca to explain the spectrum taken at $3.73$ days after the GRB with a photospheric velocity of 44,000 km s$^{-1}$. We emphasize that our first two spectra of GRB 171205A have been taken significantly earlier and the derived photospheric velocities are higher, therefore our analysis is more sensitive to the abundances in the outermost layer. Our study suggests that the existence of high-velocity Fe is directly visible as an absorption line at day 0.975 with a velocity of $\sim90,000$ km s$^{-1}$ (Fig. 2). 

In the case of GRB 161219B/SN 2016jca, a jet-like explosion with the high-velocity `Fe blob' pointing toward an observer had been suggested. A similar configuration may apply to SN 2017iuk. The jet-driven model 40A of Maeda et al.\cite{Maeda2002} predicts X($^{56}$Ni) $> 0.1$, X(Ca) $\sim$ X(Ti) $\sim 10^{-3}$, X(Si) $\sim 10^{-4}$ at $> 20,000$ km s$^{-1}$ along the jet axis. The mass fractions averaged over the whole ejecta, including the non-jet direction, are X($^{56}$Ni) $\sim$ X(Si) $\sim 0.01 - 0.1$, and X(Ca) $\sim$ X(Ti) $\sim 10^{-4}$ at $> 20,000$ km s$^{-1}$. The abundance in the outermost layer in our spectral synthesis model has a striking similarity to this prediction.  The characteristic velocity of the `jet component' in the model 40A, assuming a non-relativistic jet, is still lower than our dividing velocity between the main SN ejecta and the high-velocity component ($v_{div}\sim55,000$ km s$^{-1}$), however, this is also sensitive to the exact configuration. 
For the inner region, the same model predicts increasing mass fractions for Ca and Ti, X(Ca) $\sim 0.01$ along the jet axis and $\sim 10^{-3}$ for the whole ejecta, i.e. by a factor of 10 as compared to the outer layer. This behavior is again consistent with the spectrum-synthesis model result for SN 2017iuk, where we require an increased amount of Ca and Ti in the lower-velocity material. This analysis therefore suggests that the composition structure is well represented by a configuration containing a jet and cocoon. The jet-driven explosion produces a flat density distribution as well, as this is represented by the continuously active central engine. 

The outermost edge of the ejecta participating in the optical-spectral-line formation is located at $\sim 110,000$ km s$^{-1}$. We reject the possibility that the ejecta extends further outwards in a smooth manner towards even higher velocities, since such material should create additional absorption at even higher velocities than observed. Our rapid spectral observation and spectral modeling identify the outermost edge of the GRB-SN ejecta and such material would be hidden at later epochs without detectable traces. 

\paragraph*{The earliest spectrum at Day 0.0675}

The X-Shooter spectrum at day 0.0675, after subtracting the afterglow contribution, shows a hot continuum. Combined with the UVOT photometry the spectrum is consistent with a high-temperature blackbody. At this epoch, the photospheric velocity must be at least $110,000$ km s$^{-1}$, i.e. the outer edge participating in the later spectral formation must be opaque at this very early phase. We model the SED at this phase using analytical expressions\cite{Suzuki2017} for the blackbody emission from a relativistically expanding fireball with a constant velocity. We also assume that the photosphere is located in the outer edge of the expanding fireball. 

This is a situation very similar to the standard post shock-breakout cooling emission observed for canonical SNe, where the photosphere is supposed to be located physically in the outermost ejecta without the effect of recession of the photosphere until the temperature there has decreased below the recombination temperature\cite{Arnett1980}. While in our object the outermost layer is expected to be dominated by heavy elements, the inferred high temperature ($>100,000$ K) indicates that the outermost layer is sufficiently ionised and therefore the photosphere is supported by electron scattering. We assume that the photospheric temperature decreases as $T_{\rm ph} \propto t$ when the radiation diffusion is negligible, i.e. the behavior expected for the characteristic temperature in the adiabatically expanding fireball\cite{Arnett1982,Arnett1980}. The radiation energy content in the fireball decreases as $E_{\rm rad} \propto t^{-1}$, or equivalently $E_{\rm rad, 0}=E_{\rm rad} (t) \times (V t / R_0)^{-1}$ where $R_0$ refers the radius at the shock breakout, i.e. the progenitor radius in a canonical situation. 

Assuming $110,000$ km s$^{-1}$ as the photospheric/fireball velocity, a reasonable fit to the SED is obtained with $T_{\rm ph} \sim 300,000$ K at day 0.0675. The radiation energy content in the rest frame at day 0.0675 in this model is $\sim7\times 10^{49}$ erg, assuming spherical symmetry. If $R_0 \sim 10 R_\odot$ as expected for the WR progenitor, then the initial energy content of this component must be $\sim6\times 10^{51}$ erg. Alternatively, if the progenitor would be surrounded by a dense wind and the shock breakout would take place at $\sim 300 R_\odot$, then the initial radiation energy content in the outermost layer at the `wind shock break-out'\cite{Campana2006} will be $\sim2\times 10^{50}$ erg. These are roughly compatible with the kinetic energy content in the outermost layer in our spectral model ($\sim 10^{52}$ ergs above $\sim 55,000$ km s$^{-1}$.).

\section*{References for Methods}

\section*{\small Data Availability}

The optical spectra obtained with GTC/OSIRIS and VLT/X-shooter are available in the {\it GRBspec} repository at http://grbspec.iaa.es. The optical spectra obtained with VLT/X-shooter are also available in the WISEREP repository at https://wiserep.weizmann.ac.il/object/7496. The optical data that support the plots and the tables presented in this manuscript, as well as the python codes used for the data analysis, are available from the corresponding author upon reasonable request. The entire photometric dataset is also available in a machine readable text file available at https://osf.io/apq3d/. \textit{Swift} XRT and UVOT data are public and can be downloaded from the HEASARC archive (https://heasarc.gsfc.nasa.gov/docs/archive.html) or other \textit{Swift} archival data mirror web sites. The open-source code TARDIS used for the spectrum synthesis is available at https://tardis.readthedocs.io/en/latest/ .

\newpage

\begin{figure}
\centering
\includegraphics[scale=0.8]{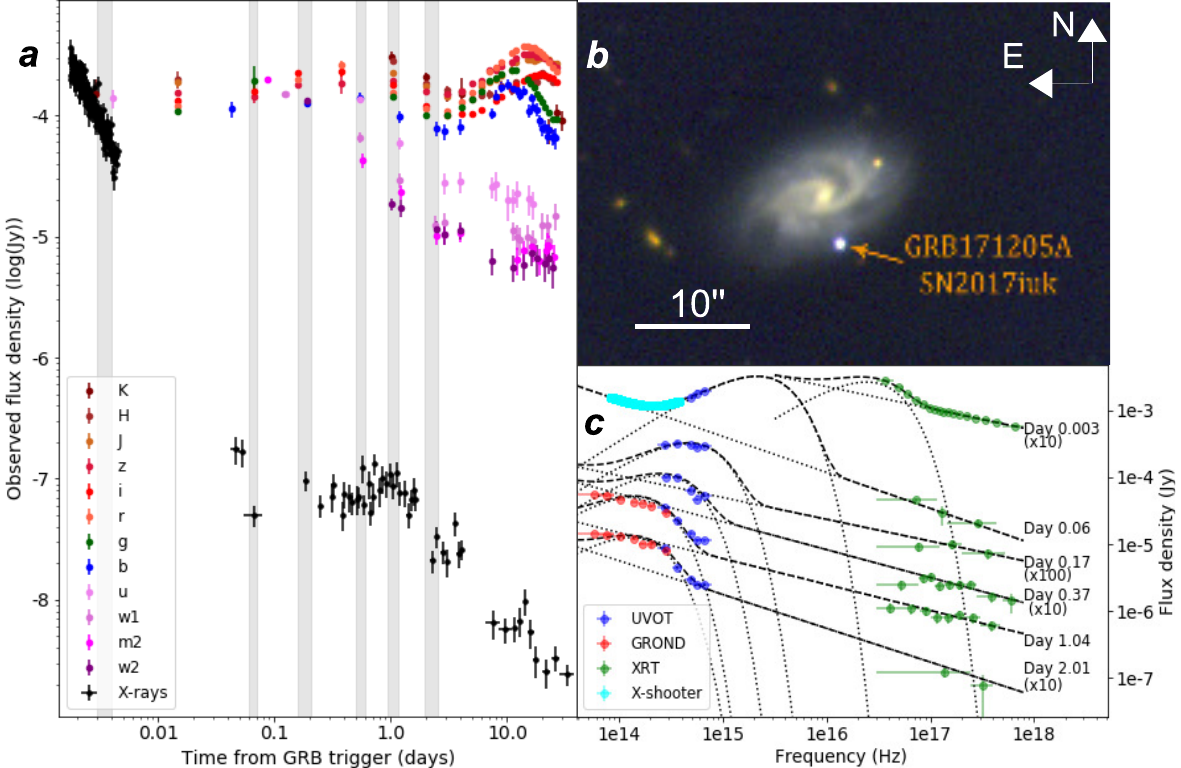}
\caption{\textit{(a)} Multi-wavelength light curve of the optical transient following GRB 171205A. During the first three days the optical, UV and NIR emission is dominated by a spectrally evolving bump. Beyond day three, a classical Type Ic supernova component emerges, SN 2017iuk. Grey regions correspond to the epochs of our multi-frequency SED analysis.  \textit{(b)} True-colour image of GRB\,171205A/SN 2017iuk and its spiral host galaxy obtained with OSIRIS at the 10.4m GTC.  \textit{(c)} The best-fit results using one or two blackbodies plus a power-law spectral model for the six spectral energy distributions, obtained with {\it Swift} for UV and X-rays and ground telescopes (GROND and X-shooter), that are marked in panel {\it (A)}.}
\label{fig:1}
\end{figure}

\newpage

\begin{figure}
\centering
\includegraphics[scale=0.6]{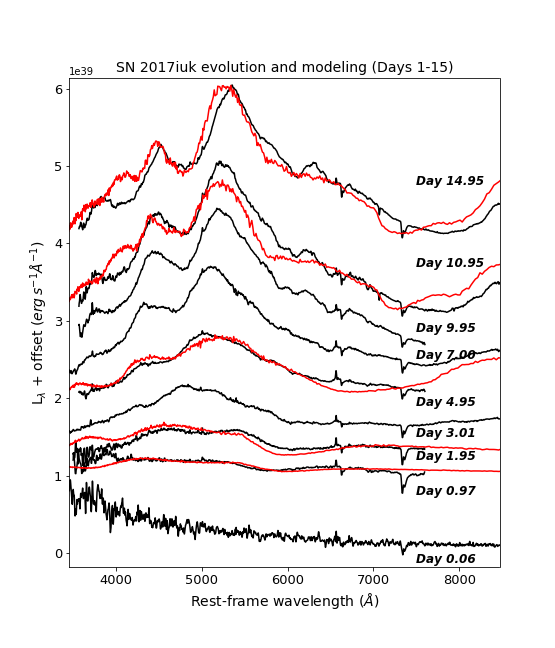}
\caption{The spectral evolution of SN 2017iuk during the first 15 days after the GRB. All spectra are shown as black curves, and they have been de-reddened for Galactic extinction, with the GRB afterglow contribution being subtracted. The simulated emission (red curves) obtained from our synthesis model for some selected spectra are shown as red curves. For the spectral simulation at Day 0.957 an arbitrary constant has been considered, due to the uncertainty in the afterglow component continuum, to match the observed data.}
\label{fig:2}
\end{figure}

\newpage
 
\begin{figure}
\centering
\includegraphics[scale=0.27]{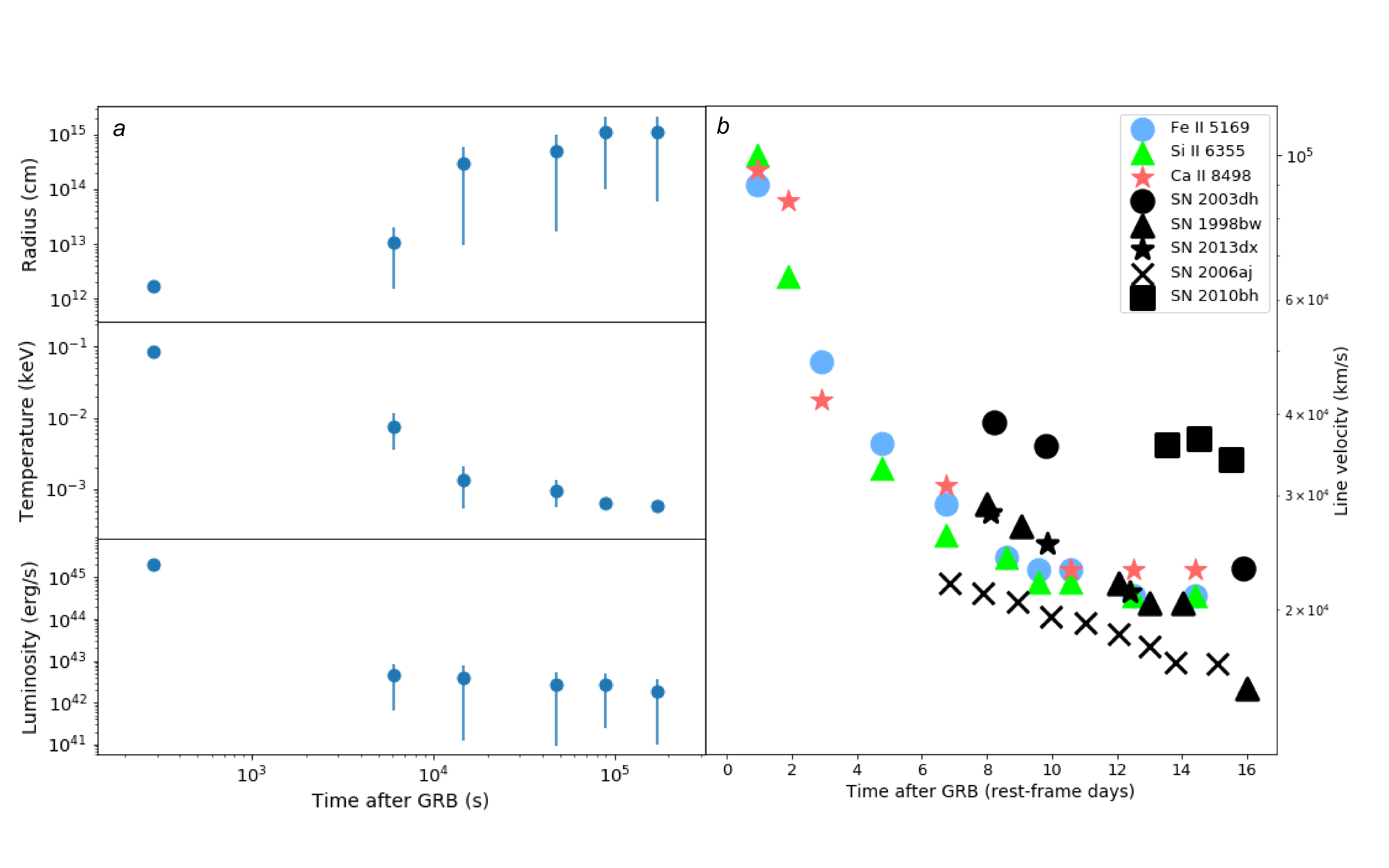}
\caption{\textit{(a)} Temporal evolution of the blackbody model parameters: from the top: radius, temperature and luminosity of the blackbody as a function of time since the GRB detection. \textit{(b)} The velocity variation estimated from the absorption line of Fe{\sc II} $\lambda$5169, Si{\sc II} $\lambda$6355 and Ca{\sc II} $\lambda$8498 in the SN 2017iuk spectra up to Day 16. For comparison we show the velocity estimated from the absorption of Si{\sc II} $\lambda$6355 for the following GRB-SNe: GRB 980425/SN 1998bw\cite{Galama1998}, GRB 030329/SN 2003dh\cite{Hjorth2003}, GRB 060218/SN 2006aj\cite{Modjaz2006}, GRB 100316D/SN 2010dh\cite{Bufano2012} and GRB 130427A/SN 2013dx\cite{Xu2013}.}
\label{fig:3}
\end{figure}

\newpage

\begin{figure}
\centering
\includegraphics[scale=0.95]{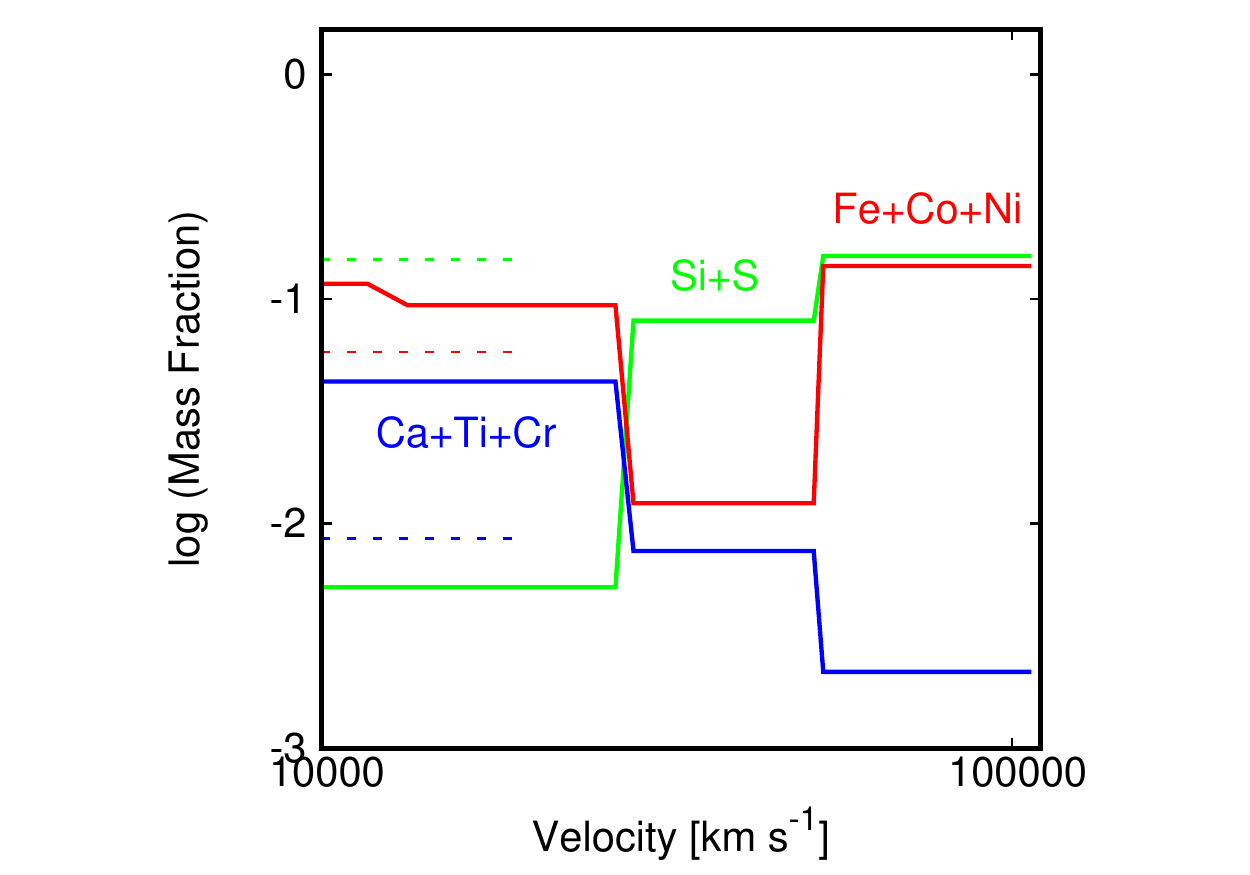}
\caption{The abundance composition structure of the SN 2017iuk ejecta that is used in our spectral synthesis model. Dashed lines represent the homogeneous mixing levels of the explosive elements. At very high velocity we observe an increase in the abundance of Fe/Ni, which is expected when the GRB jet injects metal-rich blobs over a time-scale longer than the typical shock/jet breakout.}
\label{fig:4}
\end{figure}

\newpage

\begin{table*}[h!]
\centering
\caption*{\textbf{Table 1.}\label{tab:10}
Log of the spectroscopic observations. The values in the second column refer to days after the GRB discovery. In the fifth column, the exposure times for the OSIRIS instrument refers to the grism R1000B if there is a single value, and to the R1000B and R2500I grisms where two values are given.}
\begin{tabular}{l  c c c c}     
\hline
Date & T - T$_0$ & Time & Instrument & Exp. time\\
 & (days) & (UT) & & (s)\\
\hline
2017 12 05 & 0.0625 & 08:56:18	& X-shooter	& 1x600 \\
2017 12 06 & 0.975 & 06:44:42 & OSIRIS & 2x600 \\
2017 12 07 & 1.947 & 06:05:03 & OSIRIS & 2x600 \\
2017 12 09 & 3.018 & 07:46:23 & X-shooter & 2x400 \\
2017 12 09 & 3.943 & 05:58:28 & OSIRIS	& 5x600 \\ 
2017 12 10 & 4.954 & 06:14:12 & OSIRIS & 3x600 \\		
2017 12 12 & 7.005 & 07:24:06 & X-shooter & 2x400 \\
2017 12 13 & 7.982 & 06:54:38 & FORS	& 1x600\\
2017 12 14 & 8.905 & 05:03:45 & OSIRIS & 3x600 \\
2017 12 15 & 9.947 & 06:05:03 & OSIRIS & 3x600 \\
2017 12 16 & 10.952 & 06:11:45 & OSIRIS	& 2x400+2x400 \\
2017 12 18 & 12.973 & 06:41:08 & OSIRIS	& 2x300+2x600 \\
2017 12 20 & 14.936 & 05:48:39 & OSIRIS	& 2x300+2x500 \\
2017 12 23 & 17.945 & 06:02:12 & OSIRIS & 2x300+2x500 \\
2017 12 25 & 19.850 & 03:44:34 & OSIRIS	& 2x300+2x500 \\
2017 12 26 & 21.027 & 07:59:58 & X-shooter & 2x400 \\
2018 01 01 & 26.861 & 04:00:38 & OSIRIS	& 2x300+2x500 \\
2018 01 04 & 29.926 & 05:34:52 & OSIRIS	& 2x500+2x500 \\
\hline
\end{tabular}
\end{table*}

\begin{table*}
\centering
\caption*{\textbf{Table 2.}
Fit results of the spectral energy distributions built from GROND, \textit{Swift} UVOT, and XRT data.}
\label{tab:SED}
\begin{tabular}{l  c c c c c}     
\hline
Epoch & $\Gamma$ & $kT$ & $R_{BB}$ & $L_{BB}$ & $\chi^2/DOF$\\
(days) & & (eV) & (cm) & (erg s$^{-1}$) &  \\
\hline
0.003 & 1.65$\pm$0.03 & 86$\pm$5 & (1.5$\pm$0.1)$\times 10^{12}$ & (2.1$\pm$0.2)$\times$10$^{45}$ & 904.7/765\\
0.06 & 2.07$\pm$0.22 & 7$\pm$4 & (1.0$\pm$0.8)$\times 10^{13}$ & (4.6$\pm$3.9)$\times$10$^{42}$ & 710615.8/55388\\
0.17 & 1.65$\pm$0.39 & 1.3$\pm$0.8 & (3.0$\pm$2.9)$\times 10^{14}$ & (3.9$\pm$3.7)$\times$10$^{42}$ & 138.0/31\\
0.55 & 1.83$\pm$0.19 & 0.9$\pm$0.4 & (5.0$\pm$4.8)$\times 10^{14}$ & (2.7$\pm$2.6)$\times$10$^{42}$ & 76.3/75\\
1.04 & 1.77$\pm$0.15 & 0.6$\pm$0.1 & (1.1$\pm$0.6)$\times 10^{15}$ & (2.7$\pm$2.4)$\times$10$^{42}$ & 3776.9/2128\\
2.01 & 1.88$\pm$0.60 & 0.5$\pm$0.1 & (1.1$\pm$0.8)$\times 10^{15}$ & (1.8$\pm$1.5)$\times$10$^{42}$ & 10513.7/2063\\
\hline
\end{tabular}
\end{table*}

\begin{table*}
\centering
\caption*{{\bf Table 3.} Model parameters: the photospheric velocity (second column), the photospheric temperature (third column) and the bolometric luminosity in Solar units (fourth column) are given for specific SN epochs from the GRB trigger (first column).}
\label{tab:11}
\begin{tabular}{lccc}
\hline
   $T-T_0$ & $V_{\rm ph}$     & $T_{\rm ph}$ & $log (L_{\rm bol}/L_{\odot})$            \\
    (days) & (km s$^{-1}$)    & (K) &             \\
\hline
     0.975 & 59,000 & 9,200 & 8.30 \\
     1.947 & 53,000 & 7,800 & 8.61 \\
     7.005 & 27,500 & 7,600 & 9.07 \\ 
     10.952 & 21,500 & 7,700 & 9.23\\
     14.936 & 15,500 & 8,200 & 9.22 \\
\hline
\end{tabular}
\end{table*}

\begin{table*}
\centering
\caption*{{\bf Table 4.}
Elemental abundances, obtained from the synthesis model, as a function of the expansion velocity.}
\label{NewA}
\begin{tabular}{lccccc}
\hline
 Elements & Velocity (km s$^{-1}$) & & & & \\
&  $<18,000$ & $18,000-25,000$ & $25,000-35,000$ & $35,000-55,000$ & $>55,000$ \\
\hline
C        & 1.0000E-02 & 1.0000E-02 & 1.0000E-02 & 1.0000E-02 & 1.0000E-02 \\
O        & 8.9382E-01 & 9.0110E-01 & 9.0082E-01 & 9.3120E-01 & 8.8892E-01 \\
Mg       & 5.0000E-02 & 5.0000E-02 & 5.0000E-02 & 5.0000E-02 & 5.0000E-02 \\
Si        & 5.0000E-03 & 5.0000E-03 & 5.0000E-03 & 5.0000E-03 & 5.0000E-03 \\
S        & 2.2000E-04 & 2.2000E-04 & 2.2000E-04 & 2.2000E-04 & 2.2000E-04 \\
Ca        & 7.9077E-03 & 7.9077E-03 & 4.1576E-03 & 1.5326E-03 & 7.8265E-04 \\
Ti        & 1.5765E-03 & 1.5765E-03 & 8.2654E-04 & 3.0154E-04 & 1.5154E-04 \\
Cr        & 5.2594E-03 & 5.2594E-03 & 2.7594E-03 & 1.0094E-03 & 5.0935E-04 \\
Fe        & 1.1870E-02 & 8.6700E-03 & 1.1870E-02 & 6.7000E-04 & 1.9870E-02 \\
Co        & 2.8177E-04 & 2.0177E-04 & 2.8177E-04 & 1.7700E-06 & 4.8177E-04 \\
Ni        & 1.4039E-02 & 1.0039E-02 & 1.4039E-02 & 3.8650E-05 & 2.4039E-02 \\
$^{56}$Ni & 5.6000E-02 & 4.0000E-02 & 5.6000E-02 & 0.0 & 9.6000E-02 \\
\hline
\end{tabular}
\end{table*}

\newpage

\begin{figure*}\label{fig:col}
\centering
\includegraphics[width=0.8\columnwidth]{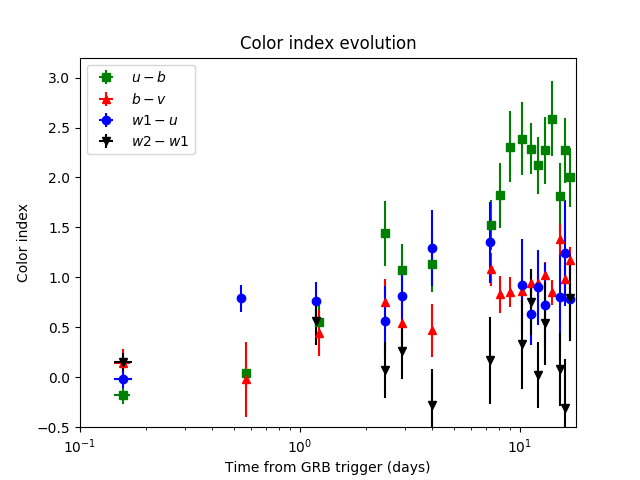}
\caption*{{\bf Figure 5. The early colour index evolution.} The $u-b$, $b-v$, $uvw1-u$ and $uvw2-uvw1$ colour index evolution, computed from UVOT data in the first 18 days after the GRB trigger. Error bars represent 1 s.d..}
\end{figure*}

\begin{figure*}\label{fig:SED}
\centering
\includegraphics[angle=0,scale=0.9]{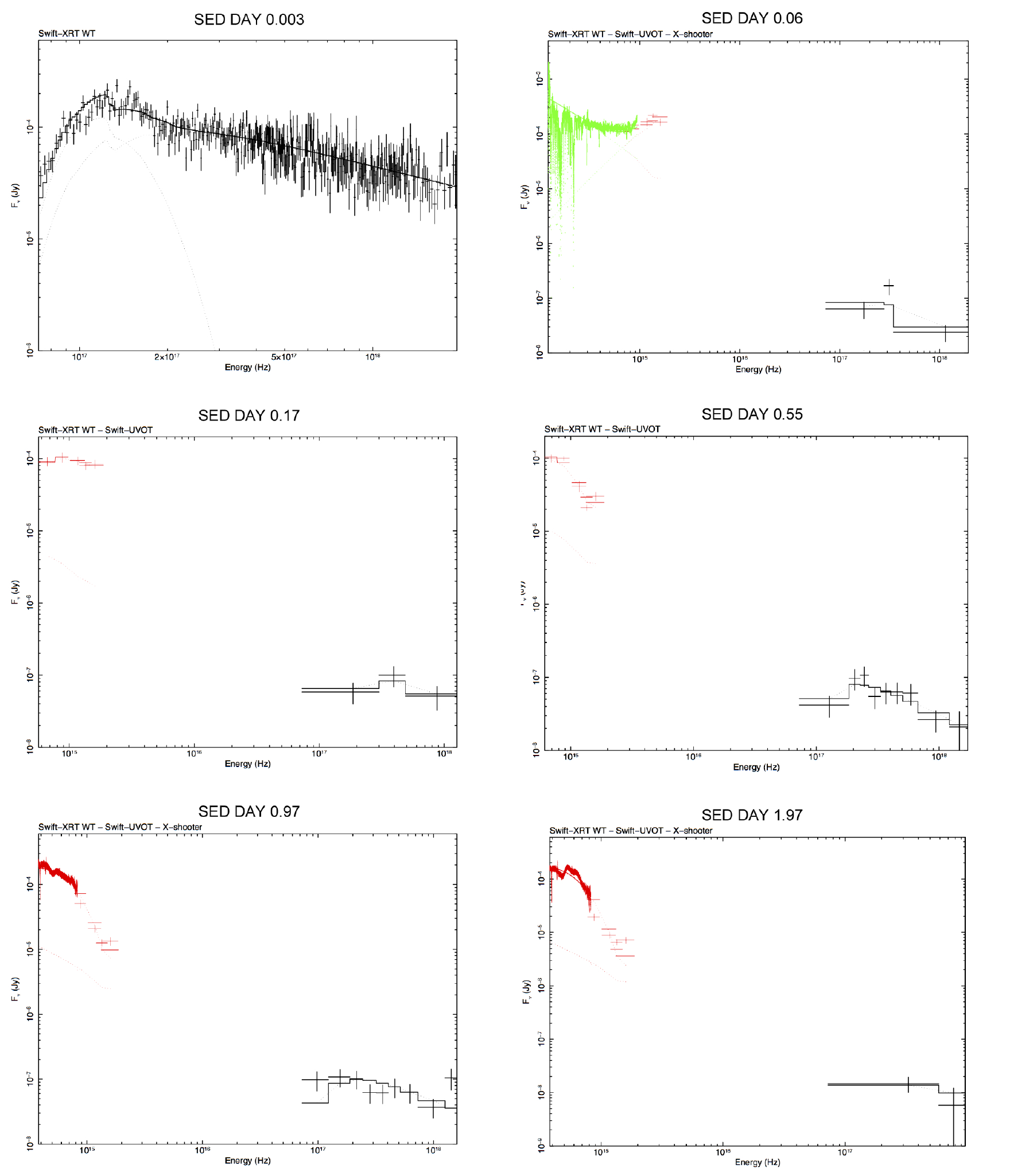}
\caption*{{\bf Figure 6.
The modelling of the spectral energy distributions.} SEDs for the epochs at {\it (top)} $T_{SED1}=0.003$ days and $T_{SED2}=0.06$ days, {\it (middle)} $T_{SED3}=0.17$ days and $T_{SED4}=0.55$ days, {\it (bottom)} $T_{SED5}=0.97$ days and $T_{SED6}=1.97$ days. All data sets use photometric data points obtained with \textit{Swift} UVOT for the low-energy part of the spectrum. Error bars represent 1 s.d. The SED is complemented with VLT/X-shooter and GTC/OSIRIS spectra for the epochs at days 0.06, 0.97 and 1.97, respectively, while for the X-ray energy range we built specific \textit{Swift} XRT spectra. An additional spectrum is shown in the top-left panel together with the best-fit results obtained for the \textit{Swift} windowed timing mode spectrum computed at 0.003 days using a blackbody + power-law spectral model.}
\end{figure*}

\begin{figure*}
\centering
\label{fig:6}
\includegraphics[width=0.9\columnwidth]{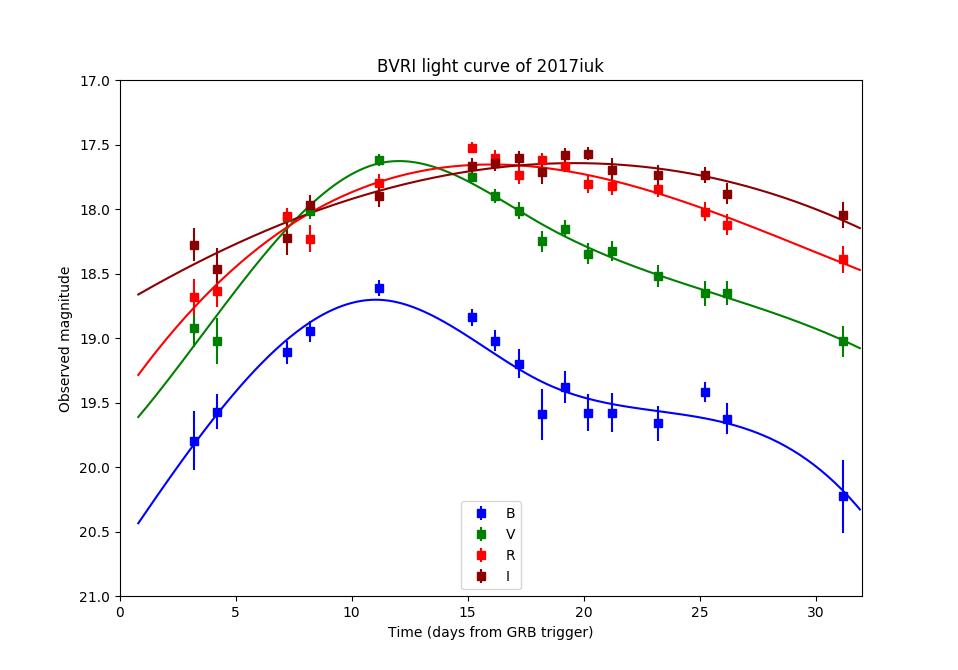}
\caption*{{\bf Figure 7.
SN 2017iuk light curve evolution.} $BVR_C I_C$ magnitude evolution of SN 2017iuk as observed with the RBT/PST2 telescope. Coloured curves represent the interpolation functions used for estimating the peak brightness. Error bars represent 1 s.d.}
\end{figure*}

\begin{figure*}
\centering
\label{fig:7}
\includegraphics[width=0.9\columnwidth]{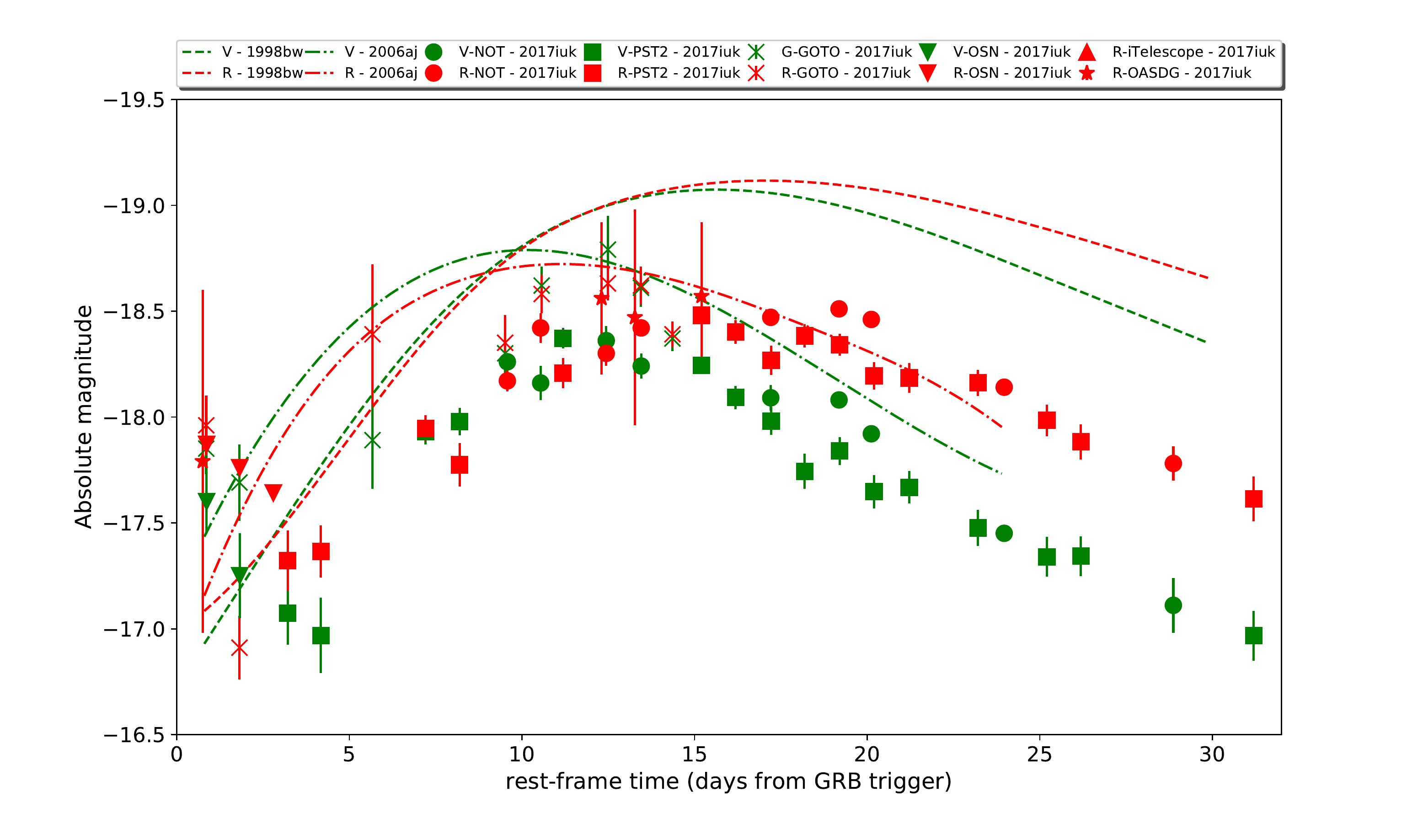}
\caption*{{\bf Figure 8.
SN 2017iuk vs. SN 1998bw and SN 2006aj.} Evolution of the $V$ (green) and $R_C$ (red) absolute magnitudes for SN 2017iuk as observed from the NOT, the OSN, the RBT/PST2, the GOTO and smaller telescopes (iTelescope, OASDG). The evolution in the first 30 days of SN 1998bw (dashed curves) and SN 2006aj (dot-dashed curves) are also shown, considering a common rest-frame time interval. Error bars represent 1 s.d..}
\end{figure*}

\begin{figure*}\label{fig:Spec1}
\centering
\includegraphics[width=0.8\columnwidth]{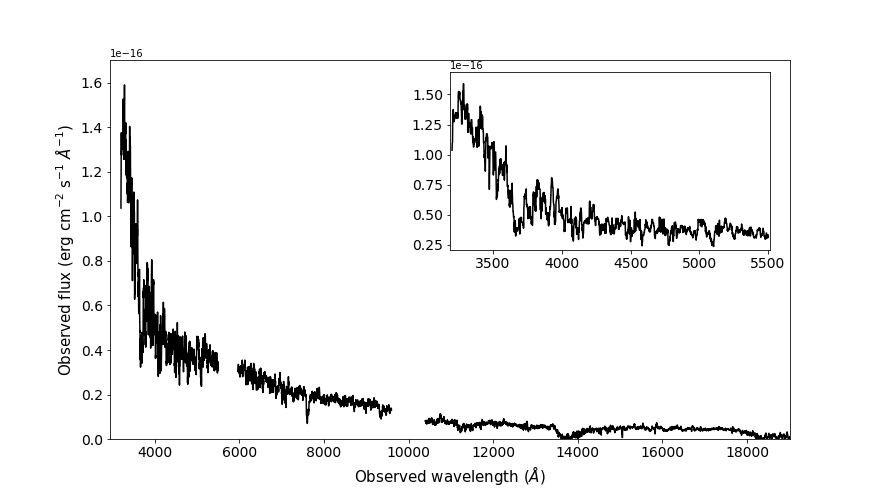}
\caption*{{\bf Figure 9.
The spectrum of GRB 171205A/SN 2017iuk obtained 1.5 hours after the GRB detection.} This spectrum has been obtained with VLT/X-shooter in the range $3200-19000$ \AA. The inset shows the UVB arm ($3200-5500$ \AA), where the emission excess at wavelengths $\leq4000$\,\AA{} and a possible absorption feature at $\sim$3700\,\AA{} are shown.}
\end{figure*}

\begin{figure*}\label{fig:He}
\centering
\includegraphics[width=0.8\columnwidth]{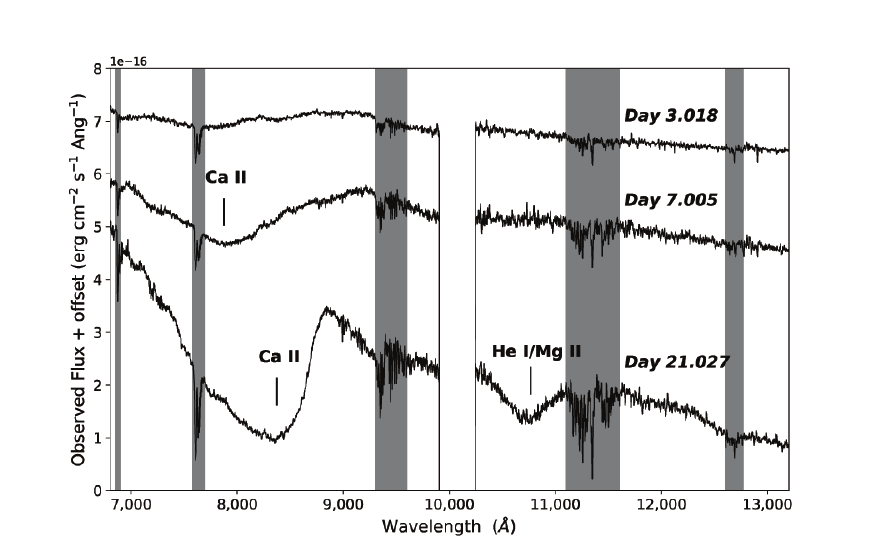}
\caption*{{\bf Figure 10.
Spectroscopic evolution of SN 2017iuk in the near-IR wavelength range.} Gray regions indicate telluric features in the spectra. The possible He{\sc I} $\lambda$10830 / Mg{\sc II} $\lambda$10914 feature is visible in the day 21 spectrum, while the Ca{\sc II} triplet shows a P-Cygni profile at bluer wavelengths.}
\end{figure*}

\begin{figure}\label{fig:KM}
\centering
\includegraphics[width=0.99\columnwidth]{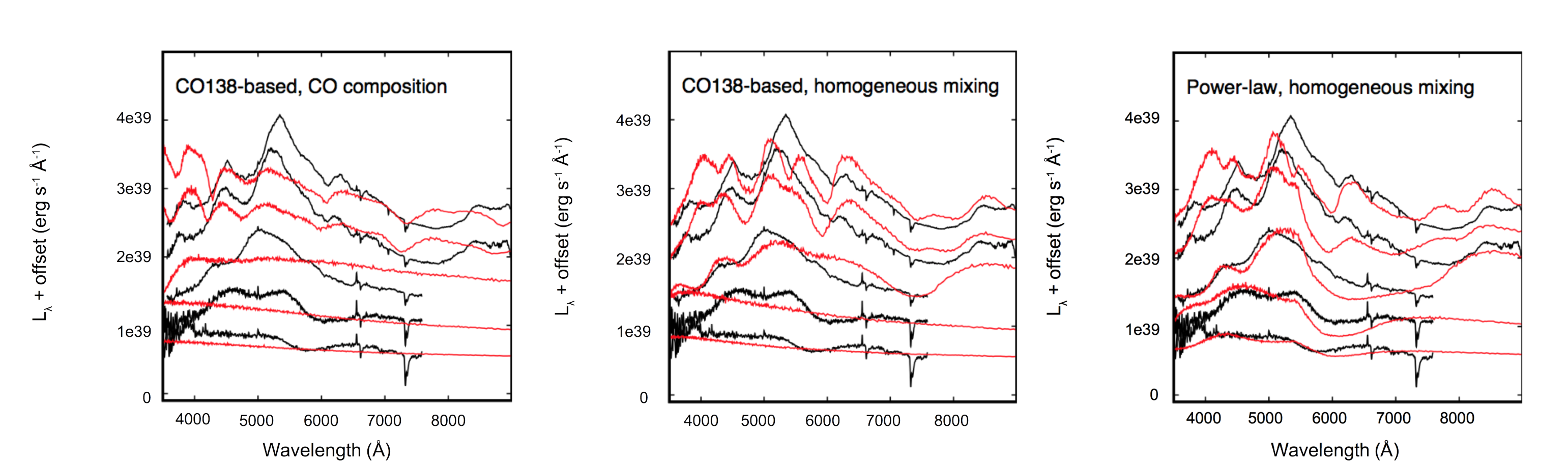}
\caption*{{\bf Figure 11.} The observed spectral sequence of SN 2017iuk from Day 0.947 to Day 14.995 (red curves) as compared with spectral synthesis simulation (black curves) in the case of a C+O core abundance \textit{(left panel)}, in the case of a homogeneous mixing of the explosive elements\cite{Nakamura2001} \textit{(middle panel)} and in the case of a homogeneously-mixed power-law distribution \textit{(right panel)}. The differences with the results using the model considered in this work (see Fig. 2) are clearly evident.}
\end{figure}

\end{document}